# Examining Requirements Change Rework Effort: A Study

Bee Bee Chua[1] and June Verner [2]
University of Technology Sydney, Australia [1]
Sydney, New South Wales
Australia 2007
bbchua@it.uts.edu.au
University of New South Wales[2]
Sydney, New South Wales
Australia 2007

## ABSTRACT

Although software managers are generally good at new project estimation, their experience of scheduling rework tends to be poor. Inconsistent or incorrect effort estimation can increase the risk that the completion time for a project will be problematic. To continually alter software maintenance schedules during software maintenance is a daunting task. Our proposed framework, validated in a case study confirms that the variables resulting from requirements changes suffer from a number of problems, e.g., the coding used, end user involvement and user documentation. Our results clearly show a significant impact on rework effort as a result of unexpected errors that correlate with 1) weak characteristics and attributes as described in the program's source lines of code, especially in data declarations and data statements, 2) lack of communication between developers and users on a change effects, and 3) unavailability of user documentation. To keep rework effort under control, new criteria in change request forms are proposed. These criteria are shown in a proposed framework; the more case studies that are validated, the more reliable the result will be in determining the outcome of effort rework estimation.

## 1. INTRODUCTION

Software maintenance is becoming a core focus in today's Information Technology business context. More companies are focusing on upgrading existing applications than on implementing new projects. A global economic downturn has unfortunately pressured many companies into withholding new projects and deferring their implementation.

Keeping project budget cost and time aligned for on time project delivery during the maintenance of an existing software project is never an easy task for a project manager. While much of the project funding is spent on requirement analysis, design, coding and testing, the remaining funding may not be enough to provide support for other software maintenance issues. Sometimes, the funding may run out completely because the effort rework risks for the cost of requirement changes have not been anticipated.

Because making requirements changes can be expensive, estimating the cost of effort rework on each change can consequently also be costly. To understand what requirements changes are needed

49

to correct or modify the system, lessons learned from previous effort estimates for requirements change rework need to be applied by IT practitioners. A requirements change can cause a ripple effect on other changes; thus an investigation into the effects of software rework is necessary.

The motivation for this paper is to provide an overview of a conceptual change management framework, empirically validated via a case study, in which a new development is found to be significantly impacted by the effect of effort rework. Further to this finding, an insight is provided into evaluating the criteria used in current change request forms and the introduction of suggested new criteria on which change forms should be based. In section 2, related software maintenance work is discussed. The research method is discussed in section 3. Results from the case study are included in sections 4 and 5. In section 6, the new criteria for inclusion in change request forms are introduced. Section 7 presents a brief discussion of the refinement of our framework, and a conclusion and outline for future work is discussed in section 8.

## 2. RELATED WORKS

The definition of a requirements change originates from the area of software maintenance and change management; each requirements change identifies the type of change, the functionality required by the change, and the effect and impact of the change.

Edelstein [1] quotes a definition of software maintenance based on IEEE standard 1219 [2], which states that it is the modification of a software product after delivery, to correct faults, to improve performance or other attributes, or to adapt the product to a modified environment. This definition is similar to the definition of the need for requirements change. Other software maintainers [3,4,5] define software maintenance by focusing their views specifically on bug-fixing, user support and system adaptation. Requirements changes as documented in change control forms provide limited information for software practitioners when they need to approve and implement the change.

The reasons for requirements changes mostly relate to error detection and correction, modification of the original requirements, changes for operational purposes and the support of user requests. Requirements changes can be categorized by type, by volume, by case study context, by domain, by change management and by their own characteristics and attributes [6-13].

User change requests and software change are the most significant reported kinds of requirements change. At the user level, user change is a typical type of requirements change. Most user requirements changes concern the analysis and design of a system. After the prototyping stage and requirement elicitation stage, users' original requirements are modified to suit their needs and to accommodate their demands for enhanced functionality and design change. These changes can be large, small, simple or complex, important or trivial, depending on the demands made by users.

The other common type of requirements change is a software change. A study of software development changes conducted by Weiss and Basili [14] of projects at the NASA Software Engineering Laboratory reported that the most frequent type of software change is an unplanned designed modification. They found that the reasons for software changes are 1) to improve program optimization, 2) to improve the services offered to its users and 3) to clarify and improve the maintainability of the software product.

A systematic literature review [2-8,13-15] revealed that many requirements change risks were found in the disciplines of project management, change management, software maintenance, information systems and software engineering and mostly related to 1) environmental-issues and 2) learning-issues. No doubt there are many other factors influencing requirements changes; however, these two groups are the largest groups critical to projects and frequently mentioned in literature.

There are four types of software maintenance: Corrective, Adaptive, Perfective and Preventive [16]. Corrective change refers to modifications initiated because of defects in the software. A defect can be an error in design, logic and coding. Adaptive change, however, is driven by the need to accommodate modifications in the environment of the software system. Perfective changes are changes undertaken in order to expand the existing requirements of a system. Preventive change is undertaken to prevent future malfunctions or to improve maintainability of the software [17- 20].

If we do not know the maintenance effort cost of making these requirements changes, no matter which category they fall into, few changes can be successfully implemented. When assessing the cost of each requirements change, effort must be taken into account. The conventional approach to estimation, which drives most project managers and software maintainers, is to apply an analogy-based approach where some value is assigned for rework effort. Modern parametric estimation tools, for example, SLIM [21] and COOCOMO 2.0 [22], can provide good estimates for software development effort and support costs. Unfortunately, these tools do not deal well with rework effort for maintenance with respect to requirements changes.

Project size estimates can be based on components, the count of function points, expert judgement, other non-parametric methods and non-algorithmic models. For many years, in all cases, expert judgment was the predominant choice for effort estimation [23].

Earlier work by [24] found that the most accurate estimates were based on analogy and expert opinion. Molokken and Jorgensen [25] make similar claims based on a review of a number of surveys carried out on software effort estimation. They find that experience-based estimation incorporated with fuzzy logic outperformed other types of techniques. Hoch et al [26] in their study on decision support systems also suggest that experts perform better than models in a predictive environment.

There are, however, some problems with effort and cost estimation. According to Putman [21] and Boehm [22], there are several drawbacks. Firstly, there is usually no good logic or rationale used to develop estimates. Secondly, there is no stable of requirements, design, coding, development processes, and no realistic interpretation of the original requirements and resource estimates from which to develop the system. Thirdly, a large number of faults have been discovered in software productivity rate estimates. Fourthly, estimation done early is less knowledgeable regarding the software to be developed which results in many estimation errors [26].

The objectives of the paper are as follows: Firstly, to address significant attributes that affect inappropriate rework effort estimation. Secondly, to propose new criteria to be included in change request forms for reviews by the change management committee or configuration team to reduce rework impact. Thirdly, to highlight the steps in our framework in order that other researchers can replicate our work with further case studies.

## 3. RESEARCH METHOD FOR FRAMEWORK VALIDATION

Case studies are specifically designed for investigative research in field studies for phenomena when 1) a large variety of factors and relationships are present, 2) no basic laws exist to determine which factors and relationships are important, and 3) when the factors of interest and their relationships can be directly observed [27].

The main emphasis of this research is on a case that concerns understanding the relationship between changes in requirements and effort rework in a change management environment. It is not a research case about organizational change and people change. It is also not an interview-based case, because it does not rely heavily on interviewing practitioners for data collection. It is an instrumental and observational case study in which the author is the primary data gatherer who observes and collects data from an IT Change Management component of an organization. It is an instrumental case because it refers to a case to gain understanding of related issues to the case. The subjects of the observational case study are IT practitioners responsible for estimating requirements changes. Because change management is only found in the software industry environment, gaining access to company data is difficult. The authorization of the organization is required together with the cooperation of management.

Records of requirements changes that have been documented and updated in change management databases for this organization mostly report on maintenance support for existing applications. The examination of IT change request forms is typical of the kind of information needed for our data collection. An overview of our research method applied in a case study context is illustrated as follows:

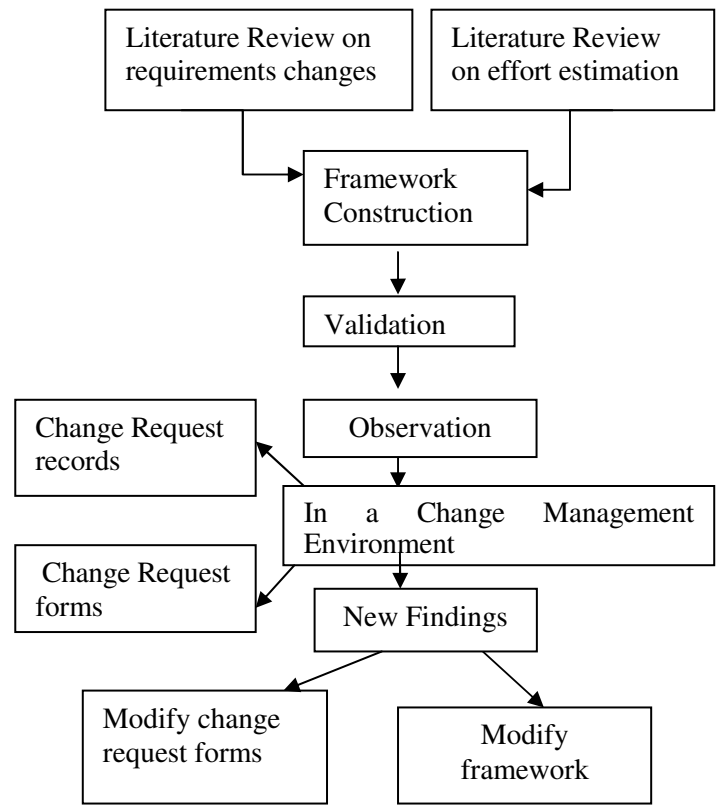

We reviewed scholarly papers on both requirement changes and effort estimation in order to understand what had previously been discussed, and what had not. Our research questions formed the basis for the development of a conceptual framework to estimate the cost of requirement changes. To evaluate the framework's applicability and reliability, we needed to test it in a change management environment. Change request records and change request forms are the primary focus of unit analysis identified in our case studies. Nonetheless, the other objective of validating the framework was not only to collect results from tests, but also to analyze results, conclude and report on new findings, with the significant aim of determining possible reasons as to why effort increases. While analyzing these results, the framework on each step was carefully reviewed by authors to seek for its consistency and accuracy. Hence, the decision on what to modify in the framework is based on data analysis of its accuracy and completeness.

## 4. FRAMEWORK TESTED IN A CASE STUDY

The aim of developing a change management framework is to guide practitioners in their control or avoidance of excessive costs when dealing with requirements change rework effort. There are other change management frameworks; however, their focus is mainly on procedures, process and people. In addition, the current parametric estimation models do not provide facilities for estimating the cost of a requirements change and or rework effort. Our conceptual framework focuses on six steps, which are reasonably easy to understand, and can guide practitioners. Previous work [12, 29] outlines the six steps of the framework (see below).

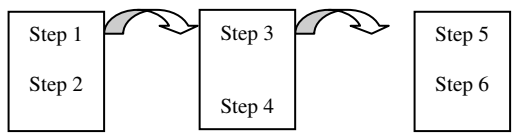

Step 1: Categorize requirements changes (RC) into first order change and second order change.

Step 2: Note the reasons for the RC.

Step 3: Understand the factors relating to, and impacting on the RC.

Step 4: Distinguish vertical and horizontal dimensions of RC relationships**.**

Steps 5 and 6: Identify the relationship between effort and various change types, and estimate the amount of person effort required. (To be incorporated with COCOMO 2.0)

The preliminary framework was validated in a medium-size software organization specializing in maintaining large in-house embedded systems. A change management system was developed for capturing maintenance issues with respect to requirements changes for large applications that are old; for example, having an application age of more than 8 years.

Incomplete and inconsistent change management records are common in any change management database so it is necessary to conduct a record filtration process. It took two person-days to complete the filtration process for 106 records. We successfully filtered 17 unwanted records that were duplicated, incomplete, or were change records that did not belong to the four maintenance types. Next, we grouped requirements changes by type to find out where they belonged. We successfully validated step 1 of the framework, which is the first order change. Because we wanted to find out which requirement changes have high effect on effort rework, our next step was to sort them by mode of function. This confirmed framework step 2 as appropriately validated. In validating step 3 of the framework, we met a challenging problem, namely, that there was insufficient information from data sets to guide us in understanding the impact and effect of requirements changes. We realized that there was a need to improve step 3 into more refined steps. The following sections explain the results collected from the case study in detail.

## 5. RESULTS FROM CASE STUDY

We applied our framework to test 106 data sets. It is clear with 89 out of 106 change records addressing adaptive and corrective changes that these are the most frequent changes reported by users.

Table 1 shows that changes related to adaptive changes and corrective changes are responsible for the submission of a higher number of change request forms than perfective and preventive changes. Interestingly these changes deal with more updates than insertions or deletions; i.e. 56 defects (14 from adaptive, 42 from corrective). Step 1 of our framework is validated successfully and the following table illustrates the findings.

| Framework step 1 | | | | | | | |
|---|---|---|---|---|---|---|---|
| First order change | | Second order of change | | | | | |
| Change Types | Total number of Change Request Records for application size ranging from 12,000 to 100,000 | Insertion | Updates | Deletion | Insert and update | Insert and delete | Update and Delete |
| Adaptive | 14 | 4 | 7 | 0 | 3 | 0 | 0 |
| Corrective | 42 | 4 | 35 | 1 | 1 | 1 | 1 |
| Perfective | 8 | 4 | 2 | 2 | 0 | 0 | 0 |
| Preventive | 6 | 3 | 2 | 1 | 0 | 0 | 0 |
| Functional Change | 19 | 14 | 4 | 0 | 0 | 1 | 0 |

**Table 1 shows statistics of change types based on the number of change request forms submitted.**

To determine change type and the reasons for change, in order to validate step 2 of our framework, we grouped the records into five kinds of change types. We followed the literature in defining each of the four maintenance types: adaptive, corrective, perfective, preventive and functional change. The following diagram (A) shows the reasons for change based on the change type.

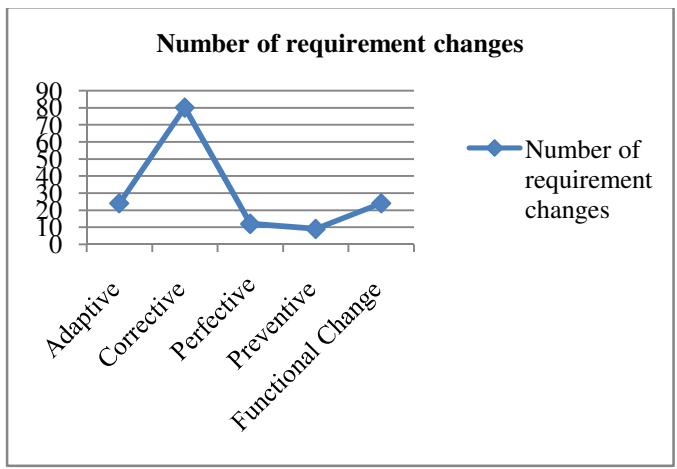

**Diagram A shows the categorization of change types by the number of change forms submitted for application size source of lines of code ranging from 12,000 to 100,000.**

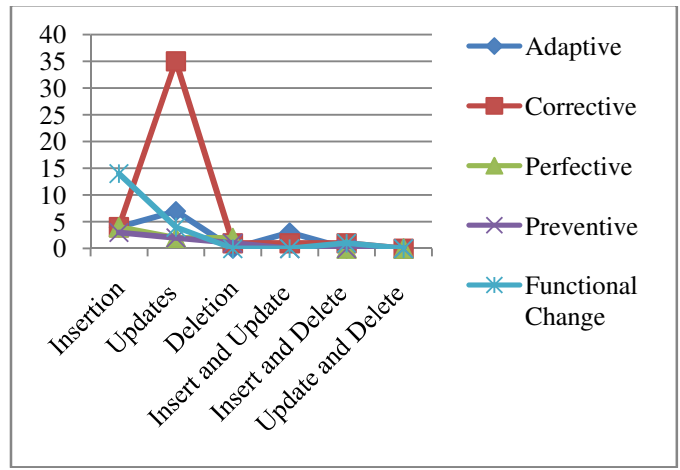

**Diagram B illustrates the function mode for each change type**

Diagram B illustrates the function mode for each change type. Typically, corrective changes have a high number of updates and this correlates to the users' requirement that defects in their applications are fixed when updating the lines of source code in the program. Ironically, a high number of change requests submitted by users with regard to functional change relate to insertion. In other words, a new feature needs to be added so that it is possible to insert a new module or modify an existing module.

From these test results we confirmed that no refinement of step 1 and 2 is required. We proceeded to validate the remaining steps shown in the framework. In the validation process for step 3 and 4 of the framework, we need to understand the factors relating to, and affecting the requirements changes; the other step distinguishes the vertical and horizontal relationships between these dimensions.

Identifying factors that are commonly discussed in literature are not our primary goal for this validation, but it is reasonable to consider whether or not they exist. Many authors discuss how business, technology, environment, people and organizational policy factors impact on a change. We were less concerned with such information, because there has been very little recent work in this area. Our approach is to investigate implicit factors not discussed widely from the perspective of change effort in the literature. We reviewed adaptive, corrective and functional changes because of their more data was found to contribute to our findings than from preventive and perfective. We discuss each change type and the result of each change in the following sections.

## 5.1 ADAPTIVE CHANGE RELEASES SORTED BY CHANGE TYPES AND REWORK EFFORT

Adaptive changes are mostly changes that relate to modification in the environment of the software system. There are 14 records of adaptive changes, of which 4 change records concern insertion, 7 change records concern updates, and 3 change records concern insert and update (see diagram C below). Note that a requirement change can be classified further according to its characteristics: A single characteristic, two characteristics, or three characteristics. A single characteristic can refer to insert, update, or delete. Two characteristics refer to a combination of 1) insert and update or 2) update and delete, and three characteristics are a combination of insert, update and delete.

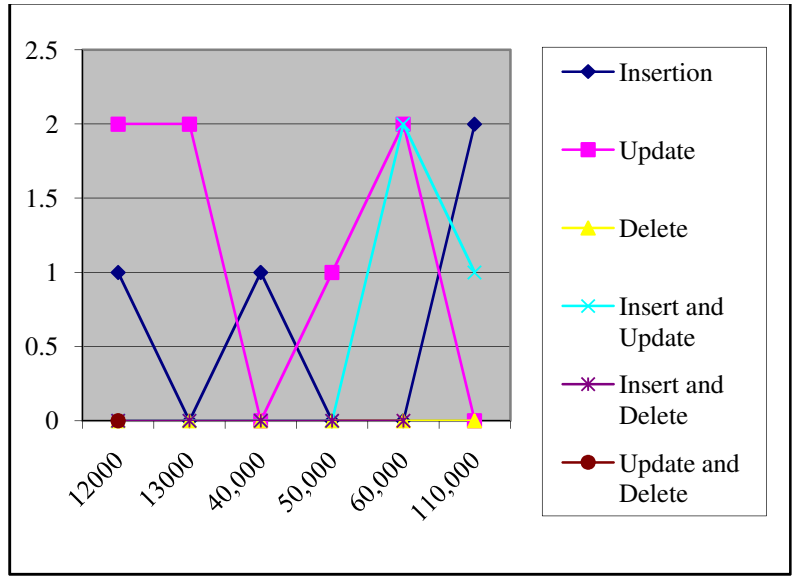

**Diagram C shows adaptive change categorized by sub-change type**

We interviewed project managers about how they measure effort and effort reworks. They told us that they assign a value to determine whether the effort is low, medium or high. If the subject has qualitative numbers, it is necessary to convert the qualitative number into a quantitative one. For example, if software maintainer confidence level is 100%, it is assigned a value of 1; if software maintainer confidence level is 50%, it is assigned a value of 2; and if software maintainer confidence level is below 50%, it is assigned a value of 3. The following attributes are needed for estimating different modes of effort reworks.

1. Effort (low, medium and high)
2. Project (size and age)
3. Requirement Change (change type and characteristic)
4. People (skill, knowledge, experience, involvement and participation in projects, and confidence level)

Four out of fourteen adaptive change records involve high effort reworks. Two are found to be from an application which has 13,000 lines of source code, and the other two are from 110 thousand lines of source code. Six changes were classified as low effort and the other four records classified were medium effort.

The software applications in this organization are rather stable, but when it comes to maintaining systems, they are program dependent. Many applications are developed in a third level of programming language and therefore the number of lines of programming code increases. Diagram D shows application size categorized by three effort types.

| Application Size | Low effort | Medium Effort | High effort |
|---|---|---|---|
| 12000 | 1 | 2 | 0 |
| 13000 | 0 | 0 | 2 |

| 40000 | 1 | 0 | 0 |
| 50,000 | 1 | 1 | 0 |
| 60,000 | 3 | 1 | 2 |
| 110000 | 0 | 0 | 2 |

**Table 2 shows effort level categorized by application size**

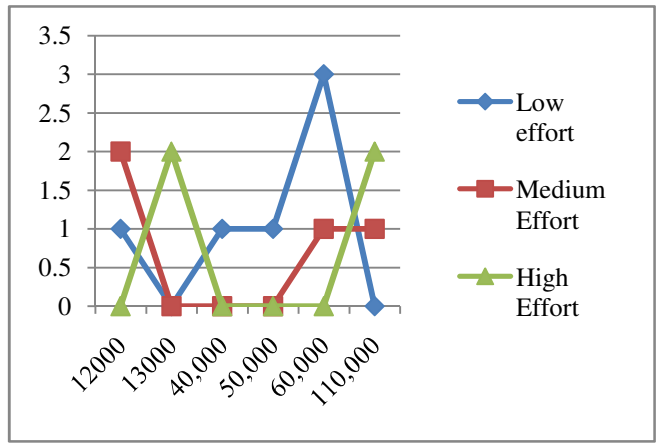

**Diagram D shows adaptive change categorized by effort levels.**

**Conclusion:** High effort rework was found in application sizes of 13,000 lines of code and 110,000 lines of source code. We reviewed the data in other columns such as programming-level (data declaration and data coding), user-level (such as user documentation and communication), system-level (unexpected problems) and system maintainers-level (for their experience, skills, knowledge, involvement and participation) and found that the problem of high rework effort can be attributed to the following variables: (1) weak characteristics and attributes in requirements, 2) poor communication by users of the change, 3) no user documentation available to understand the change, 4) unexpected problems, and 5) lack of system maintainers' effort and attributes (for example, lack of confidence and participation in the maintenance projects and maintaining tasks).

For the application size having 13,000 lines of code of requirements change to address on update, we find that there is incomplete data declaration, no user documentation, and unexpected problems. In addition, the confidence level of software maintainers is rated low in regard to solving the maintaining tasks during their participation in projects. The other example of 110,000 lines of code related to insertion has slightly different factors, because although there were no unexpected problems, the change was not discussed with users, no user documentation was provided, and software maintainers were not confident as they had not been participated in changing the main task in the project. For these reasons, we came to realise that it is important to be aware of the causes and attributes of high rework for adaptive change.

## 5.2 CORRECTIVE CHANGE RELEASES SORTED BY CHANGE TYPES AND REWORK EFFORT

According to authors [21, 22], there are several methods of measuring application size. One traditional technique is to count the function points or features in a software to determine the size of an application. Measuring the intensity of effort rework, is not dependent on the size of a requirement change; rather, it is the complexity of a requirement change that matters most in triggering effort effects. Table 3 shows the breakdown of characteristic types of corrective change by application size and the effort rework on each change for individual applications.

Diagram E shows more than one characteristic for a large size application. Between the mid-range and large range sizes of application, we notice that there are more corrective changes for single characteristics: 4 change request forms for lines of code insertion and 35 change request forms for lines of code to be updated. There is only 1 case in which the change request form has two characteristics of insert and update on a relatively small-scale application. There is no corrective change with two and three characteristics. In other words, this tells us that many of the changes found were defects in program logic.

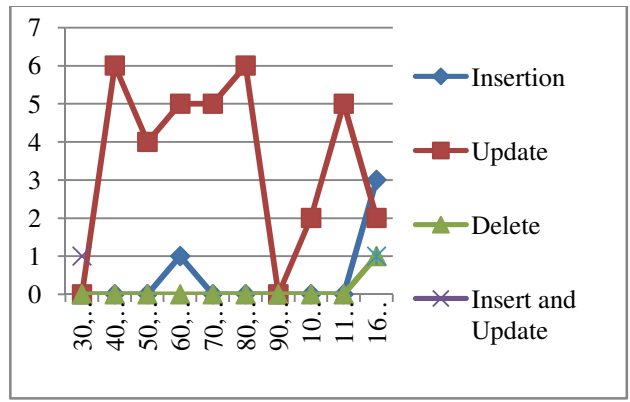

**Diagram E shows corrective change categorization**

To know which specific characteristics of change involve low and high effort rework, we present a table below that separates effort rework into different application size groupings and a diagram that shows corrective change categorization by effort types.

| Application Size | Low effort | Medium Effort | High effort |
|---|---|---|---|
| 30,000 | 0 | 2 | 0 |
| 40,000 | 0 | 6 | 0 |
| 50,000 | 0 | 3 | 1 |
| 60,000 | 1 | 5 | 0 |
| 70,000 | 0 | 5 | 0 |
| 80,000 | 0 | 6 | 0 |
| 90,000 | 0 | 0 | 0 |
| 100,000 | 0 | 2 | 0 |
| 110,000 | 4 | 1 | 0 |
| 165,000 | 3 | 2 | 1 |

**Table 3 Specific characteristics of change by effort rework**

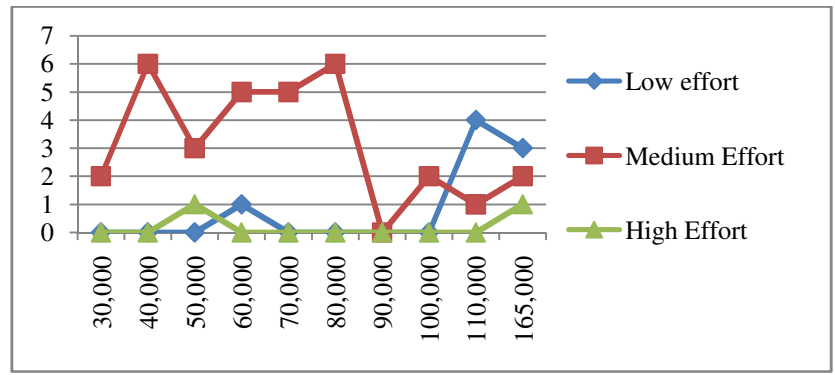

**Diagram F shows corrective change categorization by effort types**

**Conclusion**: Our analysis concludes that effort rework for corrective change is at an average level for small to large sizes of application. This information tallies with what is discussed in the literature, namely, that owing to a continuous stream of requirements changes being inserted or modified, the application becomes more complex as it is patched with numerous changes and progressively becomes more difficult to maintain [27] .

We were curious about the high rework on a particular change for an application size of 50,000. We looked into each column carefully and confirmed that the requirement for this change was 1,500 lines of source code updates at a moderate estimated rework effort of 25 person days. We compared this change with other changes having the same characteristic and noted that there was no significant difference in application age and software maintainer skill. This change required an update of 50,000 lines of source code in an application which was already 8 years old; in addition, the software maintainer had 8 years' experience of participating in this project and was confident that he knew the scope of the project well. We reviewed the data in other columns and, to our surprise, we found the high rework effort was caused by unexpected and difficult-to-understand errors. This cause is possibly correlated to 1) inconsistent or incomplete attributes in programming, 2) lack of user communication resulting in poor user involvement and failure to know of, or understand, the process of the change, 3) unavailability or complete lack of user documentation, and 4) lack of confidence by software maintainers. We never rule out that technical or systems errors are possibly resolvable, as they maybe incidental errors occurring for the first time. It is also important to bear in mind that it is highly challenging for a software maintainer to analyze a software system comprising 500,000 or 2,000,000 lines of code in order to find a hidden defect or to identify the location of a specific change of which he/she was unaware.

We want to investigate whether non-errors contribute to high effort rework and if so, the major causes of effort rework. We took data from functional change and show the results in a table and diagram in the next section.

## 5.3 FUNCTIONAL CHANGE RELEASES SORTED BY CHANGE TYPES AND REWORK EFFORT

For every application size, there is at least one functional change request from a user. This means there is no one system that can be judged as having the best feature or best functionality. Nonetheless, a reliable system is determined by the least number of errors found.  Quite a number of change requests refer to line insertions on an application size of 70,000 during that period. There is minimal functional change on update and delete. This implies that the quality of the source code is good. The diagram below shows functional change releases sorted by change type

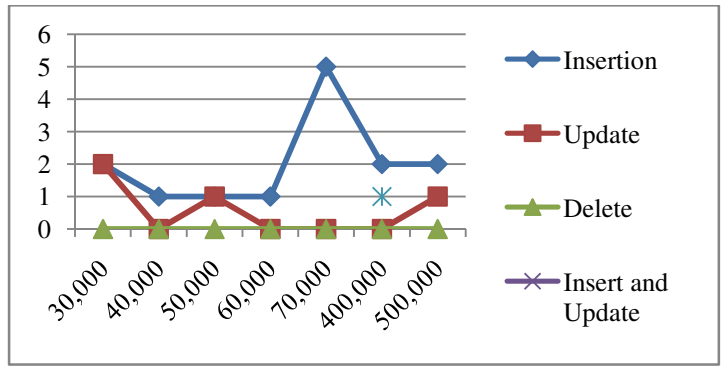

**Diagram G shows functional change releases sorted by change type**

A general view of rework effort involving functional change is that the level of this effort is usually medium to high. It tells us that although there are non-errors in normal circumstances, more rework effort applies when inserting lines of codes becomes complicated. Table 4 and diagram H shows functional change releases by effort type.

| Application Size | Low Effort | Medium Effort | High effort |
|---|---|---|---|
| 30,000 | 0 | 2 | 0 |
| 40,000 | 0 | 1 | 1 |
| 50,000 | 0 | 1 | 0 |
| 60,000 | 1 | 0 | 0 |
| 70,000 | 0 | 3 | 2 |
| 400,000 | 0 | 1 | 0 |
| 500,000 | 0 | 1 | 2 |

**Table 4 functional change releases by effort type**

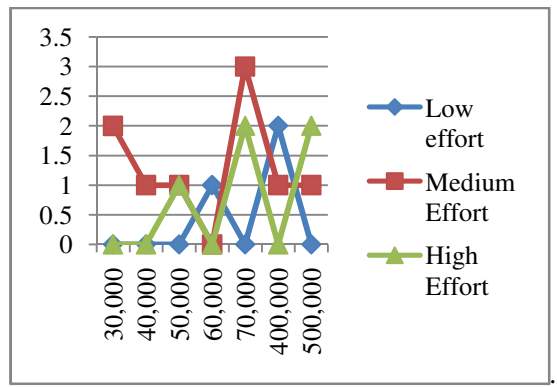

**Diagram H functional change releases sorted by effort type**

**Conclusion**: The three typical changes show high rework on application sizes of 50,000, 70,000 and 500,000. These changes that we reviewed have created new modules. For an application size of 50,000, 1,500 lines of code were inserted with 50 person days required to complete, 1,800 lines of source code inserted required 30 person days to complete for an application size of 70,000, and 1,000 lines of source code inserted required 7 person days to complete for an application size of 500,000.

Two of the maintenance requests have communicated and engaged in user participation, so we cannot conclude that the high level of rework is due to poor user communication. We confirmed that the high rework was definitely due to unexpected errors and the failure to understand them. This correlates to the weakness shown by 1) the characteristics and attributes of the lines of source code inserted, and 2) the lack of user documentation.

We did not see software maintainers' confidence level dropping as a result of dealing with functional change. All were confident they knew what to change based on the release provided. They did not find it difficult, and knew what to do based on their skill and knowledge.

## 5.4 SCENARIO CREATED TO UNDERSTAND IMPLICIT FACTORS THAT AFFECT A REQUIREMENT CHANGE

In order to successfully conduct this validation, the scenario described below helped to address implicit impact factors that may affect a requirements change and rework effort. Updates consisted of 35 corrective changes and 3 were especially selected based on the filtration sequence that we conducted. It is important that no assumptions should be allowed when carrying out this test. We have information on the features of software maintainers regarding their years of experience, their involvement in projects, the number of maintainers, project characteristics and change characteristics. As we want to focus closely on impacts, all change records must fulfil the basic criteria. These basic criteria serve the need to identify weaknesses in requirements change that can result in reworks being affected; for example, the need for questions or guidelines to users or developers while changes are corrected. Unfortunately, only three projects meet the stated conditions.

Table 5 shows the three projects named A, B and C. Each has the same number of total lines of source code but different effort rework was required for the updates. In particular, project B appears to have required more rework effort than projects A and C, the reasons for which are unexpected errors. We investigated whether this correlates to inconsistent, incorrect of data structures and data statements, and the lack of user documentation that can trigger rework. Project A has good data statements, but a lack of user involvement or user documentation can generate unexpected errors. In exploring the effects of unexpected errors that cause enormous rework, we believe that reference should be made to the paper documentation for a requirement change. Table 2 shows that project A has good data statements that unexpected errors occurred due to weak characteristics and attributes in the lines of sources code, the lack of user involvement and user documentation.

**Scenario 1 for corrective changes**

| Criteria Checks | |
|---|---|
| **People characteristics** | |
| Software maintainers' skills and experience | Same |
| Number of software maintainers involved | Same |
| **Project Characteristics** | |
| Project size | Same |
| **Change Characteristics** | |
| Update | Yes |
| Number of lines of source code to be updated | Same |
| Effort | Varies |

| | Lines of codes | '000 | | |
|---|---|---|---|---|
| | Project | A | B | C |
| Criteria | Project size | 50 | 50 | 50 |
| | Effort | 1 | 10 | 3 |
| | Lines of source code needing update | 500 | 500 | 500 |
| | New module | X | X | X |
| | Interface | X | √ | X |
| | Code declaration | √ | X | X |
| | Statement declaration | √ | √ | √ |
| | Communication with users and developers | X | X | √ |
| | User documentation | X | X | √ |
| | Unexpected errors | √ | √ | X |

**Table 5 shows three projects of corrective changes**

## 6. DESIGNING NEW CRITERIA FOR CHANGE REQUEST FORMS

We reviewed all change request forms dealing with corrective changes and discovered that the criteria on the forms provided little convincing information. The change request forms were designed to keep questions simple for end users to understand. In the proposed design of a change request form, the focus on criteria is important, particularly at the PUSSM (Programming, User, System and Software Maintainer) level.

---

**An example of new criteria in Change Request Form**

**Criteria at programming level**
Data declaration: Are they complete, accurate or consistent?
Data coding: Are they complete, accurate or consistent?

**Criteria at user level**
Documentation provided: Yes or No
Communicate this change to user: Yes or No

**Criteria at system level**
Unexpected problems: Describe the problem
Errors, explain them:

**Criteria at software maintainer level**
Skill:
Knowledge:
Experience
Involved in projects before:
How many years:
Participated in solving maintenance tasks:
How many years:

---

> **For the attention of developers**
> Please tick if you have already informed users of what you have updated, inserted, or deleted from a new module or an existing module. What would you change?

## 7. FRAMEWORK REFINEMENT

Although our framework has been validated via a case study, the evidence from our exploratory study supports the need to refine some steps in our proposed framework. Step 3, in particular, is a review of the changes of attribute types in the lines of source code for change types of an existing application, and step 4 identifies new criteria in change request forms to establish the relationship between the vertical and horizontal dimension of RC relationships**.** Steps 5 and 6 identify the relationships between effort and various change types, and estimates the amount of person effort required. The revised framework is as follows:

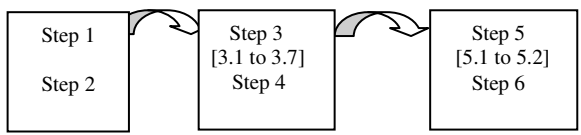

Step 1: Categorize requirements changes into first order change and second order change. (Remains unchanged)

Step 2: Note the reasons for the RC. (Remains unchanged)

Step 3: Understand the factors relating to, and impacting on the RC.

- 3.1 Know what the change type is.
- 3.2 Find out what to do with the change type. Focus on the mode of execution.
- 3.3 Check the sequence of change: update, delete or insert
- 3.4 For updates, unexpected errors that arise will add more rework. Check that variables such as attributes in data declaration and data statements are complete, correct and consistent.
- 3.5 Check user documentation is available.
- 3.6 Check the change has been communicated to users.
- 3.7 Review criteria in change request forms for any missing information.

Step 4: Distinguish vertical and horizontal dimensions of requirements change relationships**.** In this context**,** an example of a vertical dimension would be data coding or data declaration and the horizontal dimension would be referred to as inconsistent, incorrect, incomplete, missing or ambiguous.

Steps 5: Identify the relationship between effort and the various change types.

- 5.1 Review the qualifications, experience and skill of the software developers. Have they been working on projects for many years and are they aware of the change?
- 5.2 Determine the number of software developers involved in the project.

Step 6 Estimate the amount of person effort required. (To be incorporated with COCOMO 2.0 and/or other parametric models.)

## 8. CONCLUSION AND FUTURE WORK

In the literature, many researchers have focused on rework estimation problems and blame such problems on project managers and software maintainers due to their lack of experience or knowledge of estimation [22, 23, 24, 28]. They discuss the types of defects found that impact on project risk, the impact of ripple effects, improper costing and time planning, and the inability of staff to make good decisions regarding change requests [2, 7, 29].

The evidence from our exploratory study confirms that reworks of adaptive change, corrective change and functional changes are triggered by unexpected errors which correlate to 1) weak characteristics and weak attributes in data declarations or statements, 2) lack of user and developer involvement, and 3) lack of user documentation in any of the change types.

The contribution made by this paper is to provide software maintainers and change management committees with some insight into the various types of software maintenance issues which are drivers for software rework effort, using a case study to which our framework was applied. Having more detailed information can help change management committees to narrow their focus when they are considering change request forms and which mandatory criteria they require on these forms. This will enable a better understanding of the rework effort required for necessary changes and will significantly reduce the risk involved in effort estimation for the rework; better control of the changes will be possible and they will be organized more effectively and efficiently. In the future, a refinement of the framework will be possible as a result of the successful validation of other case studies.